\documentclass{aastex}
\usepackage{spr-astr-addons}
\usepackage{url}
\usepackage{longtable}
\usepackage{lscape}
\newcommand*\kps{\ensuremath{km\ s^{-1}}}

\RequirePackage{color}

\begin{document}

\title{H$_2$CO and H$_{110\alpha}$ observations towards NH$_3$ sources}
\slugcomment{Not to appear in Nonlearned J., 45.}
\shorttitle{Short article title}
\shortauthors{Autors et al.}

\author{Ye Yuan\altaffilmark{1,2}}
\affil{E-mail: yuanye@xao.ac.cn}
\and
\author{Jarken Esimbek\altaffilmark{1,3}}
\and
\author{Jian Jun Zhou\altaffilmark{1,3}}
\and
\author{Xin Di Tang\altaffilmark{1,2}}
\and
\author{Gang Wu\altaffilmark{1,2,3}}
\and
\author{Ying Xiu Ma\altaffilmark{1,3}}

\altaffiltext{1}{Xinjiang Astronomical Observatory, Chinese Academy of Sciences, Urumqi 830011, PR China.}
\altaffiltext{2}{University of the Chinese Academy of Sciences, Beijing 100080, PR China.}
\altaffiltext{3}{Key Laboratory of Radio Astronomy, Chinese Academy of Sciences, Urumqi 830011, PR China.}

\begin{abstract}
We observed the H$_2$CO ($1_{10}-1_{11}$) absorption lines and H$_{110\alpha}$ radio recombination lines (RRL) toward 180 NH$_3$ sources using the Nanshan 25-m radio telescope.
In our observation, 138 sources were found to have H$_2$CO lines and 36 have H$_{110\alpha}$  RRLs. Among the 138 detected H$_2$CO sources, 38 sources were first detected.
The detection rates of H$_2$CO have a better correlation with extinction than with background continuum radiation.
Line center velocities of H$_2$CO and NH$_3$ agree well.
The line width ratios of H$_2$CO and NH$_3$ are generally larger than unite and are similar to that of $^{13}$CO.
The correlation between column densities of H$_2$CO and extinction is better than that between NH$_3$ and extinction. These line width relation and column density relation indicate H$_2$CO is distributed on a larger scale than that of NH$_3$ and be in similar to the regions of $^{13}$CO.
Abundance ratios between NH$_3$ and H$_2$CO were found to be different in local clouds and other clouds.

\end{abstract}



\section{Introduction}
Formaldehyde (H$_2$CO) and ammonia (NH$_3$) are good tracers of the physical conditions in galactic and extragalactic molecular clouds, such as density, temperature and kinematics \citep{1993ApJS...89..123M,2013ApJ...766..108M}. Both species have been detected in a wide range of dense interstellar environments.
However, both of them suffer uncertainties caused by their formation mechanisms \citep{2006A&A...448..253L}.
It is therefore of great interest to understand how these molecules are formed, evolve, and are related in the dense region. NH$_3$ is a common used and reliable temperature probe of molecular cloud \citep{1983A&A...122..164W,2012A&A...544A.146W}. The hyperfine satellite components can give optical depths, and the population of (1,1) and (2,2) can give an indication of the kinetic temperature. NH$_3$ emissions were widely detected in molecular clouds, both in dense cores \citep{1983ApJ...266..309M} and towards ultracompact HII regions \citep{1990A&AS...83..119C}. The typical environment parameters of NH$_3$ emissions are T$\rm _K \geq$\ 10 K and n(H$_2$) $\rm \geq\ 10^4\ cm^{-3}$ \citep{1983ARA&A..21..239H}. The H$_2$CO absorption line was first detected in interstellar media by \citet{1969PhRvL..22..679S} and then collision cooling schemes of the doublet were developed \citep{1969PhRvL..22..679S,1975ApJ...200L.175G}. The scheme arrives at the lowest excitation temperature at a density region around $\rm 10^4\ cm^{-3}$ and would be quenched at the density larger than $\rm 10^6\ cm^{-3}$ \citep{1975ApJ...200L.175G}.
This density range is expected to be higher than that of $^{12}$CO and comparable with that of $^{13}$CO and NH$_3$.

Few comparisons between these two tracers were made before.
\citet{1981AJ.....86.1939S} observed NH$_3$ towards 35 strong H$_2$CO sources (the intensity of 6-cm absorption line is higher than 0.5 K) in the southern hemisphere. They reported 16 NH$_3$ detections and found no obvious statistical correlation between the intensities of the two lines.
\citet{1990ApJS...72..303N} observed H$_2$CO and NH$_3$ in five clouds for their abundance ratios to restrict chemistry models, but they did not discuss the other parameters that the two spectres can probe. The abundance ratios of the two molecules [$\rm N(NH_3)/N(H_2CO)$] they derived are mainly between 10 and $10^{3}$ in different sources.
\citet{2006A&A...448..253L} also discussed the abundance ratios of NH$_3$ and H$_2$CO in local diffuse clouds and got an abundance ratio $\rm \langle N(NH_3)/N(H_2CO)\rangle\approx0.4$.

The H$_2$CO($1_{10}-1_{11}$) transition and the NH$_3$(1,1),\\(2,2) transitions are considered to be typical high density tracers in centimeter wavelength. Theoretical calculations and observations suggested that these transitions originated under similar environment conditions, but very few works have been done to investigate the relation between them directly. So we attempted to search H$_2$CO lines towards NH$_3$ sources and to search for potential relations between the two lines. In this paper, we present a new H$_2$CO absorption and H$_{110\alpha}$ emission survey toward 180 NH$_3$ sources. We seek the relation among the H$_2$CO, H$_{110\alpha}$ and NH$_3$ lines and make an analysis as regards the H$_2$CO absorption physical conditions affected by 6-cm continuum, infrared emission and extinction.

\section{Observation and Data}

\subsection{Samples}

The sources are selected from the following papers, and the types of sources and telescopes parameters are as follows:
\begin{enumerate}
    \item \citet{1981MNRAS.195..387M}: UCHII regions, H$_2$O masers Herbig-Haro objects, reflection nebulae, CO 'hot spots' and HCN emission. \\(Chilbolton 25-m, Beamwidth: 2.2\arcmin)
    \item \citet{1993A&AS...98...51H}: IRAS sources in Orion and Cepheus molecular cloud. \\(Effelsberg 100-m, Beamwidth: 40\arcsec)
    \item \citet{1996A&A...306..267S}: Presumably massive young stellar objects, from bright 100 $\mu$m objects in IRAS Point Source Catalogue(PSC). \\(Effelsberg 100-m)
    \item \citet{2002ApJ...566..931S}: Candidates of high-mass proto-stellar objects (HMPOs), chosen from IRAS-PSC with critera: a.)detected in CS emission, b.)bright at FIR, c.)not detected in 6-cm continuum. \\(Effelsberg 100-m)
\end{enumerate}
The total number of the sources in the four papers is 218. Since our beam size of this observation is about 10\arcmin,  and to make sure the observation towards one source would not be interfered with other sources, we filtered part of the sources. Finally a list of 180 sources was used and every two sources in it are at least 10\arcmin\ apart.
Among the 180 sources, 147 were detected to have NH$_3$(1,1) emission lines in the four papers above.
The line center velocities, line widths and column densities of NH$_3$ were used in the following discussions.
The sources selected and their parameters are listed in Table \ref{table:source}, and the columns are as described in Section \ref{sec:table_source}.


\subsection{H$_2$CO and H$_{110\alpha}$ Observation}
Observations of the H$_2$CO($1_{10}-1_{11}$) $\lambda$ 6-cm absorption line ($\nu_0$=4829.6594 MHz) and H$_{110\alpha}$  RRL ($\nu_0$=4874.1570 MHz) have been made from September 2011 to August 2012, with the 25-m radio telescope of the Xinjiang Astronomical Observatory of Chinese Academy of Sciences. The telescope is located at Nanshan station (E87\degr10\arcmin40\arcsec, N43\degr28\arcmin22\arcsec). A dual-channel cooling receiver was used and provided a system temperature of $\sim$ 23 K at zenith. A Digital Filter Bank (DFB) with 8192 channels was used. 
To observe H$_2$CO and H$_{110\alpha}$ RRL simultaneously, the center frequency of the spectrometer was set at 4851.9102 MHz . A 64 MHz bandwidth was used, resulting in a velocity resolution of 0.48 \kps. Two blocks of that in DFB with the same configuration were used  simultaneously to process the two outputs of the receiver.
The half-power beam width (HPBW) of the telescope is 9\arcmin.5 at this wavelength. The DPFU (Degrees Per Flux Unit) value was 0.116 K Jy$^{-1}$ and the main beam efficiency at this wavelength is 65\%. A noise-diode was used to calibrate the flux.
The calibration of flux results in an error of about 10\%.
The observation was performed in position-switch mode.
The on-source integration time of each source is 12 minutes, but for weak sources, we added observation time to make the signal-noise ratio above 3.
\subsection{6-cm Continuum Data}
\label{sec:tc}
\emph{Sino-German 6 cm survey} \footnote{\tiny http://zmtt.bao.ac.cn/6cm/} \citep{2007A&A...463..993S} data was used as T$_C$ in Eq.\ref{eq:RT} to calculate the opacity. Of 180 selected sources 147 were located within the survey region and 33 were outside of that. For the sources not included in the survey, we observed 6-cm continuum with Nanshan 25-m radio telescope, using a cross-scan method with a bandwidth of about 400 MHz.

\subsection{Extinction Data}
\label{sec:av}
Extinction towards the sources was used to estimate the total H$_2$ column density, using the relation $\rm N(H_2)=0.94\times10^{21}A_V\ cm^{-2}$ \citep{1978ApJ...224..132B}. The extinction data calculated from 2MASS by \citet{2011PASJ...63S...1D}\footnote{\tiny http://darkclouds.u-gakugei.ac.jp/2MASS/download.html} was used.

\section{Results}
\label{sec:result}

Of the 180 sources observed, 138 sources are detected to have H$_2$CO lines and 36 to have H$_{110\alpha}$ RRL. 38 of the 138 detected H$_2$CO sources were new detections. Among the 36 RRL detections, 31 are consistent with H$_2$CO, and five sources have no H$_2$CO lines.
\label{sec:table_source}
The details are listed in Table.\ref{table:source}. The columns are as follows:
\begin{enumerate}
	\item -number
	\item -source name
	\item -right ascension of the source (J2000)	
    \item -declination of the source (J2000)
	\item -extinction of Section \ref{sec:av}
	\item -6-cm continuum brightness temperature of Section \ref{sec:tc}, in T$_{MB}$ scale
	\item -the line center velocity of NH$_3$(1,1)
	\item -the line width of NH$_3$(1,1)
	\item -column density of NH$_3$, in ($\rm 10^{13}\ cm^{-2}$)
    \item -detection flag of H110$_{\alpha}$; 0 means not detected in observation, 1 means detected
    \item -detection flag of H$_2$CO; 0 means not detected in observation, 1 means detected, 2 means detected and newly detected
	\item -reference.
\end{enumerate}

The spectral lines were reduced using CLASS software\footnote{
\tiny
CLASS is a part of GILDAS package, developed by IRAM (Institute de Radioastronomie Millim$\rm\acute{e}$trique). http://www.iram.fr/IRAMFR/GILDAS/
}. For each source, all scans were averaged. Then a polynomial baseline was subtracted and gaussian fittings were made.
Table.\ref{table:h2co} gives the fitting results. The columns are as follows: 1)-number; 2)-source name, the second, third and fourth velocity components are marked with b, c, and d if existing; 3)-LSR (Local Standard of Rest) velocity of line center and fitting error; 4)-peak intensity of H$_2$CO lines and fitting error; 5)-FWHM of line and fitting error; 6)-column density of H$_2$CO in $\rm 10^{13}\ cm^{-2}$.
For the sources with more than one velocity component, we assumed that the component the velocity of which is closest to the NH$_3$ velocity is associated with the NH$_3$ source, which was demonstrated to be reasonable by \citet{2007ApJ...668.1042K}.

The column densities of H$_2$CO were derived using the formulation in \citet{1988A&A...191..313P}:
\begin{equation}
\label{eq:colh2co}
N(H_2CO)=9.4\times10^{13}\ \tau_{peak}\cdot\Delta V\ (cm^{-2})
\end{equation}
where $\rm \tau_{peak}$ is the optical depth and $\rm \Delta V$ is the line width.
The constant $ 9.4\times10^{13}$ has already been corrected for orho-para ratio and rotational ladder for kinetic temperature at 10K.

Because the line widths of our sources are too large, the hyperfine structures (HFS) of H$_2$CO($1_{10}-1_{11}$) are blended and cannot be used to calculate the optical depth. We derived the optical depth through radiative transfer equation.
The H$_2$CO apparent optical depth $\rm \tau_{app}$ was determined through a simplified radiative transfer equation.
\begin{equation}
\label{eq:RT}
T_L=(T_{ex}-T_C-T_{CMB})\times(1-e^{-\tau_{app}})
\end{equation}
where $\rm T_L$ is the line brightness temperature, $\rm T_{ex}$ is the excitation temperature of this transition, $\rm T_C$ is the continuum brightness temperature, and $\rm T_{CMB}$ is the brightness temperature of the cosmic microwave background (CMB).
The $\rm T_L$ and $\rm T_C$ values can be obtained through our observations directly, while the $\rm T_{ex}$ values are not clear to us.
Several previous observations that can resolve the HFS of H$_2$CO reported that the $\rm T_{ex}$ values of this transition range from 1.5K to 2.0K \citep{1973ApJ...183..441H,1983A&A...118..337V,2004ApJ...614..252Y}. So we assumed the $\rm T_{ex}$ value to be the mean value of what they reported, 1.7K.
Also, we made RADEX \citep{2007A&A...468..627V} Non-LTE radiative transfer calculations to investigate the possible values of $\rm T_{ex}$. The calculations show that the $\rm T_{ex}$ tends to be lower than the range above by about 0.5K and the consequent departure in column density will be less than half an order (see Appendix \ref{sec:radex1}), comparable with the uncertainty of the orth-para ratio estimated by \citet{2012ApJ...749L..33D}.
Besides, the $\rm T_{kin}$ used in the RADEX calculation is the temperature of NH$_3$, which is the temperature of the center of the core and usually is higher than the average value across the core. So we may have overestimated the efficiency of the collision. The departures of $\rm T_{ex}$ and column density, indeed, should be smaller than the values estimated above.

\begin{figure}[t]
   \centering
   \includegraphics[width=0.9\hsize]{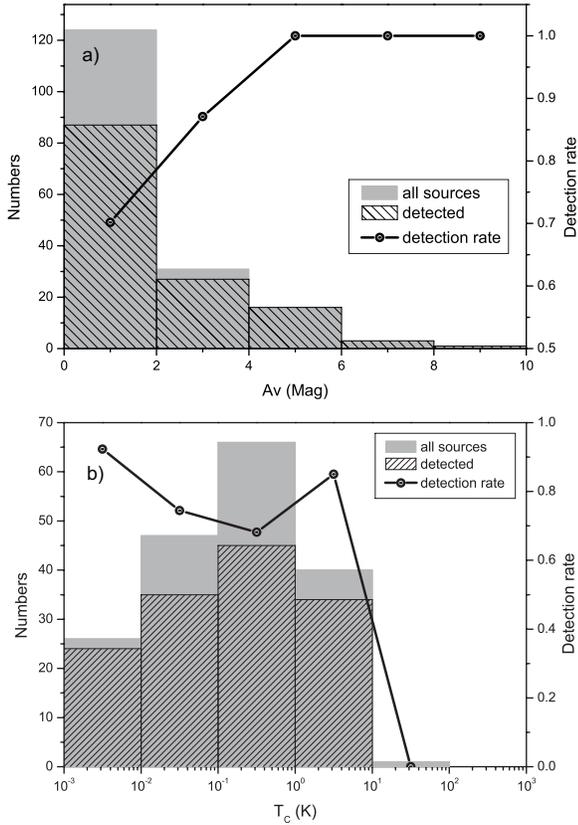}
   \caption{Detection rate distribution of H$_2$CO. The \emph{gray histogram} is the distribution of all the 180 selected sources in each a) extinction interval, b) continuum emission temperature interval. The \emph{shaded histogram} is the distribution of sources detected H$_2$CO. The \emph{solid line} is the detection rate, using the right axis.}
    \label{fig:det_av}%
\end{figure}

\section{Discussions}
\subsection{Influence of Extinction and Continuum Emission}
The H$_2$CO detection rate of different extinction and continuum intervals are plotted in Fig.\ref{fig:det_av}. The detection rate monotonically increases with the increasing of extinction. The sources with extinction larger than 4 mag have 100\%\ detection rate. All the 42 sources that did not detect H$_2$CO have the feature of low extinction.
This suggests that the H$_2$CO absorption associates well with the dense region of the molecular clouds. Comparative studies of the H$_2$CO and $^{12}$CO showed that the 4.8-GHz H$_2$CO absorption line is similar to the $^{12}$CO, tracing the lower density region of the molecular cloud \citep{2006AJ....131.2921R,2012Ap&SS.337..283Z,2013A&A...551A..28T}.
This may indicate that the H$_2$CO absorption line can still probe the molecular cloud well, although it is believed sometimes to be dominated by background continuum radiation.

A high H$_2$CO detection rate of continuum sources (see Fig.\ 1.b)) indicates that the continuum emission strongly affects the H$_2$CO absorption line. The observation of \citet{1980A&AS...40..379D} and \citet{2012Ap&SS.337..283Z} showed that the H$_2$CO absorption line has a good correlation with the 6-cm continuum background.
Toward the continuum sources, the continuum radiation plays an important role in the determination of the H$_2$CO line intensity according to Eq.\ref{eq:RT}. Therefore, the H$_2$CO absorption line intensities depend not only on the gas density, but also on the background continuum level.

The relation between detection rate and continuum brightness temperature is not as clear as that between detection rate and extinction (see Fig.\ref{fig:det_av}.b).
One of the causes may be the different column densities of the foreground gas. Different column densities of gas may cause different optical depths and then influence the absorption.
Besides, when the gas density is not high enough, strong continuum emission can increase $\rm T_C$ and $\rm T_{ex}$ in Eq.\ref{eq:RT} simultaneously (see Appendix \ref{sec:radex1}).
So a higher $\rm T_C$ may not necessarily increase the absorption even without the disturbance from column density.
Complicated relations between $\rm T_L$ and $\rm T_C$ were also reported by \citet{2012Ap&SS.337..283Z} and \citet{2014arXiv1402.5471D}. Their results show that there is a certain correlation between $\rm T_L$ and $\rm T_C$, but with considerable scatter.

\subsection{H$_2$CO Absorption and H$_{110\alpha}$ Emission}
\begin{figure}[h]
   \centering
   \includegraphics[width=0.90\hsize]{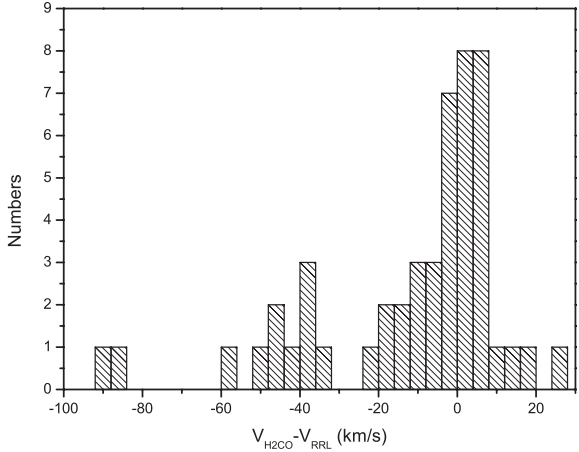}
   \caption{Histogram of the line center velocity differences between H$_2$CO and H$_{110\alpha}$. All the velocity components of H$_2$CO are used.
            }
    \label{fig:h2co-rrl}
\end{figure}

Of all the 180 observed sources 36 have observed the H$_{110\alpha}$. In all the sources where H$_{110\alpha}$ were detected, five sources have no H$_2$CO, namely the following: IRAS00494+5617, IRAS03595+5110, IRAS05327-0529, IRAS06068+2030 and IRAS18151-1208. It may be caused by the lack of the foreground dense gas in front of the HII region. Figure.\ref{fig:h2co-rrl} shows that most sources have a low absolute value (-10 $<$ V(H$_2$CO) - V(H$_{110\alpha}$) $<$ 10 km s$^{-1}$) of two lines center velocity differences, which suggests that the H$_2$CO clouds associate well with the HII region and the relative motion between H$_2$CO clouds and HII region is quite small. This result agrees with the comparative results of H$_2$CO clouds and HII region by \citet{1980A&AS...40..379D} and \citet{1987A&A...171..261C}. A few cases of H$_2$CO absorption components have large velocity differences with H$_{110\alpha}$ (above 20 km s$^{-1}$), which suggests that those H$_2$CO clouds could not be associated with a HII region. Similar results have been noted by \citet{1987A&A...171..261C}, \citet{1988A&A...191..313P}, \citet{2011A&A...532A.127D}, and \citet{2011RAA....11..156H}.

\begin{figure}[h]
   \centering
   \includegraphics[width=0.99\hsize]{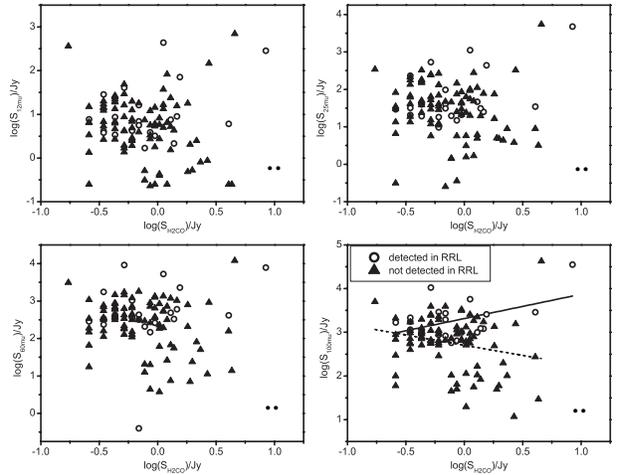}
   \caption{
   Relationship between H$_2$CO flux density and infrared flux density of: a) 12$\mu m$; b) 25$\mu m$; c) 60$\mu m$; d) 100$\mu m$.
   The \emph{cycles} are the sources detected in RRLs, and the \emph{triangles} are sources not detected in RRLs.
   The \emph{solid line} in d) is the fitting result of sources detected in RRLs, and the \emph{dash line} is that of sources not detected in RRLs.
   }
    \label{fig:flux-IR}
\end{figure}

\subsection{H$_2$CO Absorption and Infrared Emission}
There are 152 IRAS sources in our observed sample. 110 H$_2$CO absorptions and 25 H$_{110\alpha}$ emissions were detected in all IRAS sources.
A high detection ratio (about 72\%) of H$_2$CO absorption shows that the excitation of the H$_2$CO absorption line may have some connection with the infrared emission. We compared the relation between the H$_2$CO flux density and IRAS flux density in the Fig.\ref{fig:flux-IR}.
The sources were divided into two subsamples by the detection of RRLs.
The fitting between 100 $\mu$m and H$_2$CO flux density was made.
The correlation factor of sources detected in RRLs is 0.47 and the correlation factor of sources not detected in RRLs is 0.24.
This result shows that sources detected in RRLs have a certain correlation between 100 $\mu$m and H$_2$CO flux density, while sources not detected in RRLs have no obvious correlation between 100 $\mu$m and H$_2$CO flux density.
This agrees with the result of \citet{2011A&A...532A.127D}.
The 100 $\mu$m flux densities of the sources not detected in RRLs have a lower value and a larger scatter.
Without a HII region, the interstellar radiation field may contribute a lot in the 100 $\mu$m flux of the sources not detected in RRLs and make the behaviors complex \citep{2005ApJ...635L.173W}.

\subsection{Comparison between H$_2$CO and NH$_3$}

NH$_3$ is a high density tracer of $\rm n[H_2] > 10^4\ \ cm^{-3}$ \citep{1986A&A...157..207U}, while the density of H$_2$CO absorption line is considered to be $\rm 10^3<n[H_2]<10^5\ \ cm^{-3}$ \citep{1975ApJ...196..433E,1980A&A....82...41H}.
The lower density limit of H$_2$CO is little than that of NH$_3$ and nearly the same as the density at which $^{13}$CO can probe \citep[$\rm \sim 10^3\ cm^{-3}$, ][]{1997ApJ...482..245U,2012A&A...544A.146W}.
So it would be inferred that the H$_2$CO roughly traces the region between that of $^{13}$CO and NH$_3$.

\subsubsection{Line-center Velocity and Line Width}
\label{sec:lw}

The correlation of NH$_3$ and $^{13}$CO (1-0) emission towards 870$\mu m$ sources was analyzed by \citet{2012A&A...544A.146W}.
They observed NH$_3$ lines towards the APEX Telescope Large Area Survey (ATLASGAL) sources and extracted $^{13}$CO(1-0) lines of the sources from the University-Five College Radio Astronomy Observatory Galactic Ring Survey (GRS) \citep{2006ApJS..163..145J}. The line center velocity differences between $^{13}$CO(1-0) and NH$_3$ lines are smaller than the NH$_3$ (1,1) line width for most of the sources. The line width of $^{13}$CO is mostly broader than NH$_3$ lines. Turbulence or multiple structures within a large scale was considered to cause the $^{13}$CO line width to be broader than that of NH$_3$ \citep{2012A&A...544A.146W}.

\begin{figure}[t]
   \centering
   \includegraphics[width=0.99\hsize]{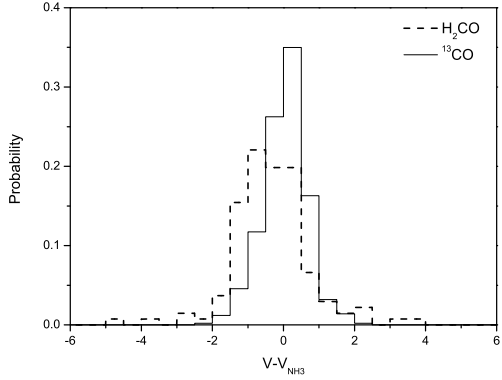}
   \caption{The probability distribution of the $\rm V_{LSR}$ difference between molecules and NH$_3$. The \emph{dashed line}is the probability distribution of the velocity difference between our H$_2$CO observation and NH$_3$; the \emph{solid line} is that between $^{13}$CO and NH$_3$ from \citet{2012A&A...544A.146W}.
            }
    \label{cum_v-v}
  \end{figure}

We analyzed the line center velocities of NH$_3$ and H$_2$CO, and we found that 90\% of the sources have a velocity difference $<$2.4 \kps and 65\% have a difference $<$1 \kps (see Fig.\ref{cum_v-v}). This difference distribution is smaller than that between $^{12}$CO and NH$_3$ shown in \citet{1990A&AS...83..119C}, which they considered to be small enough to originate in the same dynamical system.
Although more sources have negative velocity differences than positive values, this is of marginal significance given our velocity resolution.
For most of the sources, the velocity difference is smaller than both the H$_2$CO line width and the NH$_3$ line width (see Fig.\ref{cum_v-v}). This behavior is similar to the results of the observations that searched for core-to-envelope relative motions \citep{2007ApJ...655..958W,2012A&A...544A.146W}.

Both H$_2$CO and NH$_3$ have HFS, so the observed line width cannot represent the Doppler width directly, as \citet{1990ApJS...72..303N} suggested when they attempted to compare the line widths of these two lines.
Usually, fitting to a less blended HFS component is used to get the intrinsic line width in this case, such as method NH3(1,1) fitting in CLASS software. 
But for H$_2$CO ($1_{10}-1_{11}$) lines, the HSF components are too close together compared to the line width to make such a fitting. Even the farthest component from the group, F=1-0, is only about 1.2 \kps away from the group center, which can easily be blended in most sources, as is shown in \citet{1971ApJ...169..429T}.
To solve this problem, we introduced a similar simulation following \citet{1998ApJ...504..207B}, to get a simulated intrinsic line width from direct observed value.
The details of this simulation are presented in Appendix \ref{app1}, and the result shows that the difference between blended line width and intrinsic line width is less than 20\% for most of our sources. So we did not distinguish between observed line width and intrinsic line width in the following discussions.

\begin{figure}[t]
   \centering
   \includegraphics[width=0.90\hsize]{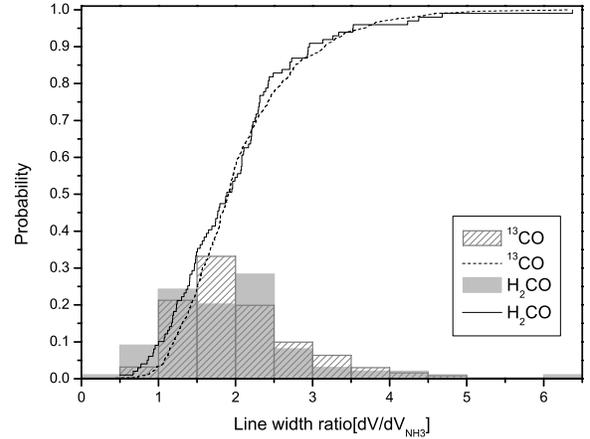}
   \caption{The probability distribution of line width ratio between molecules and NH$_3$, dV$_{^{13}CO}$/dV$_{NH_3}$ and dV$_{H_2CO}$/dV$_{NH_3}$. The \emph{solid line} and \emph{shaded histogram} are cumulative curve and probability distribution of the ratio between H$_2$CO and NH$_3$; the \emph{dashed line} and the \emph{gray histogram} are that between $^{13}$CO and NH$_3$ from \citet{2012A&A...544A.146W}.}
              \label{lwrate-cumu}%
\end{figure}

The mean value of H$_2$CO line width is 3.3\kps, and that of NH$_3$(1,1) lines is 1.9\kps.
On average, the line width of H$_2$CO is larger than that of NH$_3$(1,1).
We plotted the probability distribution of the line width ratio between H$_2$CO and NH$_3$ (dV$\rm_{H_2CO}$/dV$\rm_{NH_3}$) in Fig.\ref{lwrate-cumu}. 
The distribution of the line width ratio between $^{13}$CO (1-0) and NH$_3$ (1,1) (dV$\rm_{^{13}CO}$/dV$\rm_{NH_3}$) from \citet{2012A&A...544A.146W} was also attached to Fig.\ref{lwrate-cumu}. 
The curve shows that more than 88\% of the sources have a H$_2$CO line width which is larger than NH$_3$ line width, and 80\% of the sources have H$_2$CO line width which is 1.2 times larger than that of  NH$_3$.
The distribution of dV$\rm_{H_2CO}$/dV$\rm_{NH_3}$ is very similar to that of dV$\rm_{^{13}CO}$/dV$\rm_{NH_3}$. The average of dV$\rm_{^{13}CO}$/dV$\rm_{NH_3}$ is 2.1 and that of dV$\rm_{H_2CO}$/dV$\rm_{NH_3}$ is 1.9. The max difference (D) of the two cumulative probability curve is 0.109. The Kolmogoroff-Smirnov (K-S) statistic probability (P value) of this D value is 0.265 with the parameter N=82.72, or P=0.570 with N=50.
This similarity may indicate that the H$_2$CO line exist in the similar regions as the $^{13}$CO does, which was considered to trace a larger scale than the NH$_3$ dose and to be broader than NH$_3$ line because of turbulence within that larger scale \citep{2012A&A...544A.146W}. 
This also agrees with the spatial distribution relation between H$_2$CO and $^{13}$CO observed in \citet{2013A&A...551A..28T}.

\subsubsection{Column Density and Abundance}

\begin{figure}[h]
 \centering
 \includegraphics[width=.99\hsize]{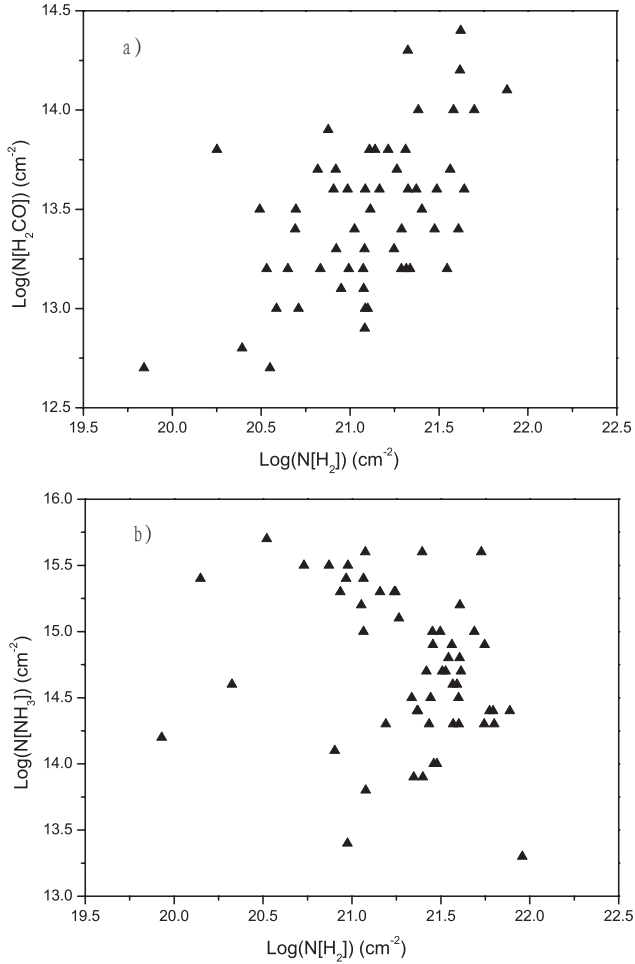}
 \caption
 {
 H$_2$CO and NH$_3$ column density plotted against H$_2$ column density. The \emph{upper panel} (a) is the column density of NH$_3$ and the \emph{lower panel} (b) is that of H$_2$CO. The H$_2$ column density was calculated using $\rm N(H_2)=0.94\times10^{21}A_V\ (cm^{-2})$ in \citet{1978ApJ...224..132B}.
 }
 \label{figcolcols}%
\end{figure}

The H$_2$CO and NH$_3$ column densities are plotted against the H$_2$ column density in Fig.\ref{figcolcols}. To match the regions observed by the molecules, we used the extinction data of the center (1\arcmin) for NH$_3$ and the average of extinction data within 10\arcmin\ for H$_2$CO.
The correlation between N[H$_2$CO] and N[H$_2$] seems to be better than that between N[NH$_3$] and N[H$_2$].
The Pearson correlation factor for data in Fig.\ref{figcolcols}.a) is 0.54 and that for Fig.\ref{figcolcols}.b) is -0.28. The extinctions are usually considered to be able to trace low H$_2$ density regions, even lower than $^{13}$CO does \citep{2009ApJ...692...91G,2006ApJ...649..807P}. Then a better correlation between H$_2$CO and a low-density-sensitive tracer indicates that the H$_2$CO may also tend to trace the low density region. This agrees with the relative position of H$_2$CO and NH$_3$ got in Section \ref{sec:lw} through the line width relation.

\begin{figure}[h]
 \centering
 \includegraphics[width=.90\hsize]{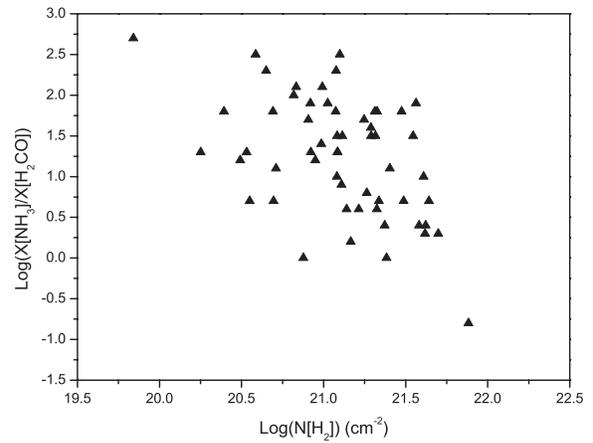}
 \caption
 {Abundance ratio of NH$_3$ and H$_2$CO vs. H$_2$ column density.}
 \label{figxx_nh2}%
\end{figure}

\begin{figure}[h]
 \centering
 \includegraphics[width=.99\hsize]{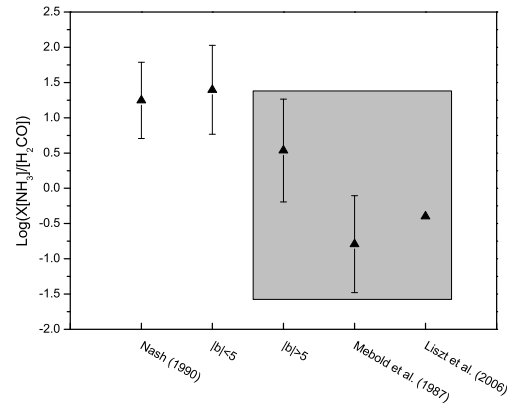}
 \caption
 {The abundance ratio of NH$_3$ and H$_2$CO in local clouds and other clouds. The five columns of data points are the average of: models in \citet{1990ApJS...72..303N}; sources in our work that have galactic latitude smaller than 5\degr; sources in our work that have galactic latitude larger than 5\degr; data of \citet{1987A&A...180..213M}; data of \citet{2006A&A...448..253L}. The error bar is the standard deviation.}
 \label{figabun_gb3}%
\end{figure}

The abundance ratio of $\rm X[NH_3]/ X[H_2CO]$ was assumed to be the ratio of the column densities following \citet{1990ApJS...72..303N}.
In the range of H$_2$ column density N(H$_2$) $\rm \sim 10^{19} - 10^{22}\ (cm^{-2})$, the majority of the abundance ratios log(X[NH$_3$]/X[H$_2$CO]) have a range from 0 to 3 and the mean value of them is 1.26 (Fig.\ref{figxx_nh2}).
This mean value is approximately equal to the ratio predicted by models in \citet{1990ApJS...72..303N}, which gives an average value of 1.25 for $\rm log(X[NH_3]/ X[H_2CO])$.
But observations of \citet{1987A&A...180..213M} and \citet{2006A&A...448..253L} showed rather different values for the abundance ratio, -0.79 and -0.40 for $\rm Log(X[NH_3]/ X[H_2CO])$ respectively.
\citet{1987A&A...180..213M} observed three high latitude clouds with these two molecules and \citet{2006A&A...448..253L} observed towards two local diffuse clouds
, while \citet{1990ApJS...72..303N} observed several sources in five clouds, and most of which have a latitude between $\rm \pm 1\degr$ .
According to \citet{2008hsf2.book..813M}, most of the molecular clouds at high galactic latitude are proximity to the Sun. Then we may reach a hypothesis that the abundance ratios of NH$_3$ and H$_2$CO are different between local clouds and other clouds.
The sources of \citet{1987A&A...180..213M} and \citet{2006A&A...448..253L} are considered to be located in the local clouds and have a lower abundance ratio.
We divided our sample into two subsamples by their galactic latitudes. The sources with $\rm |b|>5\degr$ were assumed to be local clouds.
The average value of the abundance ratio were calculated in each subsamples.
The average abundance ratios of the two subsamples and the three literatures above are plotted in Fig.\ref{figabun_gb3}
The sources suspected to be local clouds (inside the \emph{gray box}) seem to have a lower $\rm X[NH_3]/X[H_2CO]$ ratio than the others. This result suggests that there might be some differences in physical and chemical conditions between the two types of clouds.

\section{Summary}

We observed 180 NH$_3$ sources and detected 138 sources with H$_2$CO ($1_{10}-1_{11}$) lines and 36 with H$_{110\alpha}$ RRLs. Among the 138 H$_2$CO detections, 38 were considered as newly detected.
The detection rate of H$_2$CO increases monotonically with the increasing of extinction, but it seems to have a complex relation with 6-cm continuum radiation.
The line center velocity differences between H$_2$CO and NH$_3$ are small, which suggests that the two molecules should belong to the same dynamical system.
The average line width of H$_2$CO is 2.1 times that of NH$_3$, and the probability distribution curve of the line width ratios between H$_2$CO and NH$_3$ is close to that of $^{13}$CO.
This line width relation suggests that H$_2$CO may be distributed in a larger and lower density region than NH$_3$ does, and this is similar to that of $^{13}$CO.
The H$_2$CO column density correlated to H$_2$ column density (derived from extinction) well. This also suggests that the H$_2$CO exists in a larger scale than NH$_3$.
The Abundance ratio $\rm X[NH_3]/X[H_2CO]$ varies from 1 to 10$^3$ and is significantly smaller in local clouds.


\acknowledgments
We thank all the staff of Nanshan Observatory for observations.
This work was funded by The National Natural Science foundation of China under grant 10778703 and partly supported by China Ministry of Science and Technology under State Key Development Program for Basic Research (2012CB821800) and the National Natural Science foundation of China under grant 11373062, 11303081 and 10873025.

\bibliographystyle{spr-mp-nameyear-cnd}
\bibliography{yybapss-ep}

\appendix
\twocolumn
\section{Simulation of Blended H$_2$CO HFS}
\label{app1}

The optical depth of all the HFS is simply assumed to be a superposition of N components,
\begin{equation}
\label{eq:hfs}
\tau(v)=\sum_{j=1}^{N} \alpha_j\ \tau_0\ exp\Big\{-4\ln2\Big[\frac{v-(v_j+v_0)}{\Delta v_{int}}\Big]^2\Big\}
\end{equation}
where $\alpha_j$ and $v_j$ are the intensity and velocity of the \emph{j}th HFS component, and $\Delta v_{int}$ is the intrinsic line width.
The $\alpha_j$ and $v_j$ data of H$_2$CO($1_{10}-1_{11}$) HFS components were got from \citep{1971ApJ...169..429T}.
Because the optical depth of H$_2$CO($1_{10}-1_{11}$) in most cases is far less than 1, we assume that the intensity is proportional to the opacity. So the spectra line profile is the same as $\tau(v)$.

\begin{figure}[h]
\centering
\includegraphics[width=1.0\hsize]{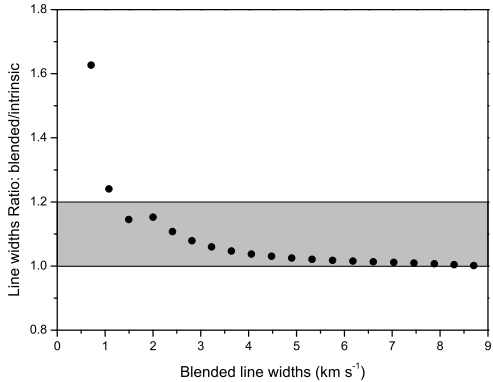}
\caption
{
The ratio of blended line width to intrinsic line width is plotted against the blended line width. The {\it gray box \rm} is the region with a ratio between 1.2 and 1.
}
\label{hfs1}
\end{figure}

Spectra for $\Delta v_{int}$ between 0.43 to 8.69 at intervals of 0.43km/s (7KHz to 140KHz at intervals of 7KHz) were calculated using Eq.\ref{eq:hfs}. Then single gaussian fittings were made to get the blended line width. The simulated spectra are shown in Fig.\ref{hfs2} and the fitting results are listed in Table. \ref{table:a1}. The ratios between blended and unblended line widths are plotted in Fig.\ref{hfs1}. Because most of our observed line widths are larger than 1\kps, this HFS blending effect would not make much difference as regards our sources.

\begin{table}[h]
\caption{HFS simulation result\label{table:a1} }             
\centering                          
\small
\begin{tabular}{c c }        
\hline\hline
Intrinsic linewidth & Observed linewidth \\
(\kps) & (\kps) \\
\hline
0.43  &  0.71  \\
0.87  &  1.08  \\
1.30  &  1.49  \\
1.74  &  2.00  \\
2.17  &  2.41  \\
2.61  &  2.81  \\
3.04  &  3.22  \\
3.48  &  3.64  \\
3.91  &  4.06  \\
4.35  &  4.48  \\
4.78  &  4.90  \\
5.21  &  5.32  \\
5.65  &  5.75  \\
6.08  &  6.18  \\
6.52  &  6.60  \\
6.95  &  7.03  \\
7.39  &  7.45  \\
7.82  &  7.88  \\
8.26  &  8.29  \\
8.69  &  8.70  \\
\hline
\end{tabular}
\end{table}


\section{Non-LTE Radiative Transfer of H$_2$CO}
\label{sec:radex1}
To absorb the CMB radiation as observed, the H$_2$CO $1_{10}-1_{11}$ transition must have an excitation temperature lower than CMB. The different cross sections of transitions in collision are considered to be the most likely reason to generate such a low  excitation temperature \citep{1969ApJ...157L.103T,1975ApJ...200L.175G,1975ApJ...196..433E}. The excitation of H$_2$CO should be treated with a non-LTE method, because the LTE approximation would cause an excitation temperature above CMB.
We used the Radex Non-LTE molecular radiative transfer code \citep{2007A&A...468..627V} to calculate the excitation of H$_2$CO.
The predicted line brightness temperature ($\rm T_{L}$) and excitation temperature ($\rm T_{ex}$) were calculated from the input H$_2$ number density ($\rm n[H_2]$), H$_2$CO column density ($\rm N[H_2CO]$), background brightness temperature ($\rm T_{bg}$) and kinetic temperature ($\rm T_{kin}$), where $\rm T_{bg}$ is the background brightness temperature, which is equal to $\rm T_C+T_{CMB}$ in Eq.\ref{eq:RT}.

\begin{figure}[h]
\centering
\includegraphics[width=0.99\hsize]{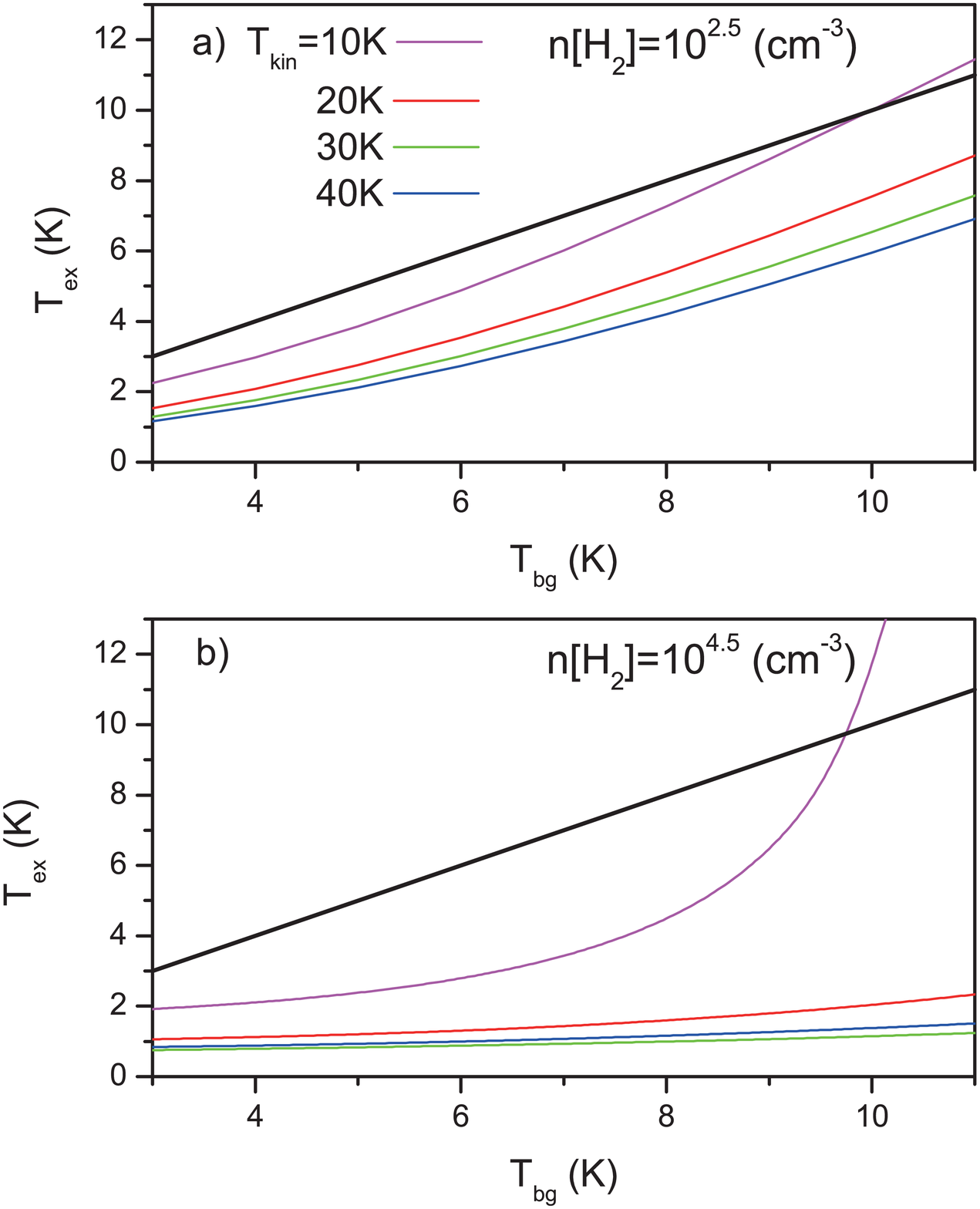}
\caption
{
Excitation temperature of H$_2$CO vs. background brightness temperature. The \emph{thick black line} is for $\rm T_{ex}= T_{bg}$.
}
\label{fig:tex1}
\end{figure}

The result shows that the competition between collision and background radiation dominates the excitation process of H$_2$CO. 
Figure.\ref{fig:tex1} shows $\rm T_{ex}$ vs. $\rm T_{bg}$ under different $\rm T_{kin}$ and n[H$_2$].
Under both low n[H$_2$] and high n[H$_2$] conditions, $\rm T_{ex}$ is less sensitive to $\rm T_{kin}$ at high $\rm T_{kin}$ ($\rm T_{kin}>20K$). 
A similar relation between $\rm T_{ex}$ and $\rm T_{kin}$ was obtained by \citet{2012ApJ...749L..33D} in high redshift galaxies.
At low n[H$_2$], collisions are insufficient. Then the background radiation dominates the level populations and the excitation temperature of H$_2$CO almost follows the variation of $\rm T_{bg}$, i.e. $\rm T_{ex}\approx T_{bg}-Constant$, as is shown in Fig.\ref{fig:tex1}.a).
This may contribute to the less clear relation between the detection rate and $\rm T_C$ in Fig.\ref{fig:det_av}.b).
At high n[H$_2$], collisions dominate the level populations and the excitation temperature keeps almost constant over the whole $\rm T_{bg}$ range.

\begin{figure}[t]
\centering
\includegraphics[width=0.99\hsize]{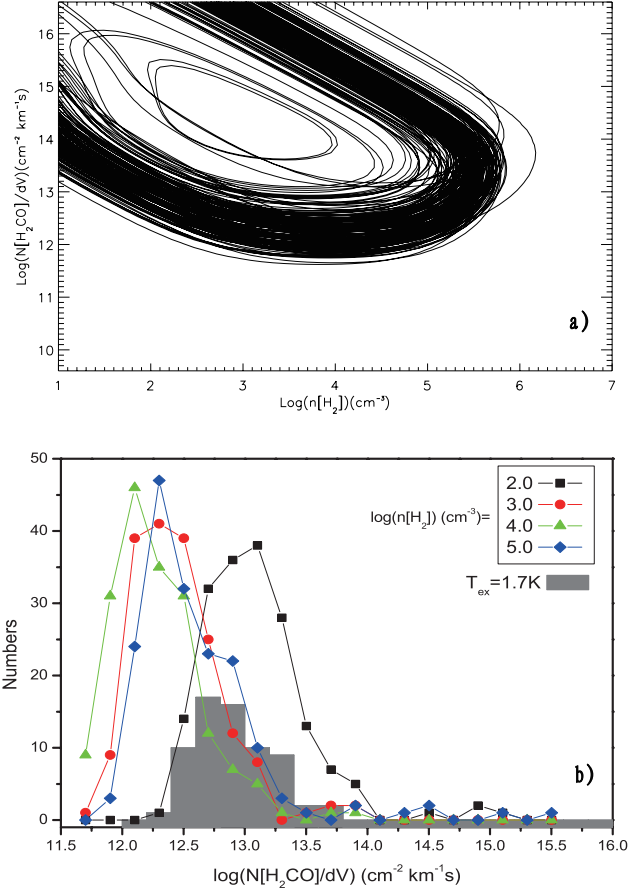}
\caption
{
a),Contours of $\rm T_{L}$ in $\rm log(n[H_2])-log(N[H_2CO])$ space. The figure presented a map of calculated $\rm T_{L}$ in $\rm log(n[H_2])-log(N[H_2CO])$ space. For each source, one contour on which $\rm T_{L}$ is equal to the observed $\rm T_{MB}$ is drawn. b) Histogram of $\rm N[H_2CO]$ under different assumptions: n[H$_2$]=$\rm 10^2,\ 10^3,\ 10^4,\ 10^5\ (cm^{-3})$ and $\rm T_{ex}$=1.7K.
}
\label{fig:tr_contr1}
\end{figure}

The possible n[H$_2$] values of H$_2$CO 6-cm lines are considered to be $\rm 10^3<n[H_2]<10^5 (cm^{-3})$ \citep{1975ApJ...196..433E,1980A&A....82...41H}.
The exact value of n[H$_2$], which in principle can be derived through the relative intensity of $1_{10}-1_{11}$ (6 cm) and $2_{11}-2_{12}$ (2 cm) \citep{1980A&A....82...41H,2008ApJ...673..832M}, is unknown because only the 6-cm lines were observed.
But in that n[H$_2$] interval, the n[H$_2$] seems not to influence the $\rm N[H_2CO]$ too much, as the contours in that region in Fig.\ref{fig:tr_contr1}.a) are almost parallel to the horizontal axis.
To find the potential uncertainty introduced by the unknown n[H$_2$], we calculated the N[H$_2$CO] under the assumptions that n[H$_2$]=$\rm 10^2,\ 10^3,\ 10^4,\ and\ 10^5\ (cm^{-3})$ respectively. 
The histogram distributions of the column densities are plotted in Fig. \ref{fig:tr_contr1}.b). The column density distributions of n[H$_2$]=$\rm 10^3,\ 10^4,\ and\ 10^5\ (cm^{-3})$ are almost the same, and that of n[H$_2$]=$\rm 10^2\ (cm^{-3})$ is about one order larger than the others.
We also plotted the histogram distribution of column densities under the assumption that $\rm T_{ex}$=1.7K in Section \ref{sec:result}.
The $\rm N[H_2CO]$ distribution of $\rm T_{ex}$=1.7K is about half an order larger than that of $\rm 10^3<n[H_2]<10^5 (cm^{-3})$.

\section{Tables}

\onecolumn
\begin{deluxetable}{cccccccccccc}

\rotate
\tabletypesize{\scriptsize}
\tablewidth{0pt}
\tablecaption{Source\label{table:source}}

\tablehead
{
\colhead{Number}                    &
\colhead{Source}                    &
\colhead{RA(J2000)}                  &
\colhead{Dec(J2000)}                 &
\colhead{Av}                        &
\colhead{T$_C$}             &
\colhead{V$_{LSR(1,1)}$}            &
\colhead{$\Delta V_{(1,1)}$}        &
\colhead{N[NH$_3$]}                 &
\colhead{H110$_{\alpha}$\tablenotemark{a}}  &
\colhead{H$_2$CO\tablenotemark{b}}  &
\colhead{ref\tablenotemark{c}} \\

\colhead{}                    &
\colhead{}                    &
\colhead{h:m:s}                  &
\colhead{\degr:\arcmin:\arcsec}                 &
\colhead{(mag)}                        &
\colhead{(K)}             &
\colhead{(\kps)}            &
\colhead{(\kps)}        &
\colhead{($10^{13}cm^{-2}$)}                 &
\colhead{Detected?}  &
\colhead{Detected?}  &
\colhead{} \\
}

\startdata
001   &  IRAS00338+6312     &   00:36:47.5        &   +63:29:02               &  1.03    &   ...     &  -17.8           & 2.3                &  99.8               &  0                   & 1         & 3    \\
002   &  IRAS00494+5617     &   00:52:23.7        &   +56:33:45               &  0.87    &   0.67    &  -30.3           & 2.6                &  87.0               &  1                   & 0         & 2    \\
003   &  IRAS01194+6136     &   01:22:48.6        &   +61:52:15               &  4.76    &   0.16    &  ...             & ...                &  ...                &  0                   & 1         & 3    \\
004   &  IRAS02219+6152     &   02:25:42.8        &   +62:06:05               &  3.20    &   8.84    &  ...             & ...                &  ...                &  0                   & 1         & 3    \\
005   &  IRAS02230+6202     &   02:26:50.8        &   +62:15:52               &  1.49    &   3.70    &  ...             & ...                &  ...                &  1                   & 1         & 3    \\
006   &  IRAS02232+6138     &   02:27:01.1        &   +61:52:14               &  2.20    &   1.20    &  -46.5           & 2.7                &  51.4               &  1                   & 1         & 3    \\
007   &  IRAS02244+6117     &   02:28:14.0        &   +61:31:12               &  1.33    &   0.91    &  -49.7           & 2.1                &  ...                &  1                   & 1         & 3    \\
\enddata

\tablenotetext{a}{Flag 0 means not detected; 1 means detected}
\tablenotetext{b}{Flag 0 means not detected; 1 means detected; 2 means newly detected}
\tablenotetext{c}{Column ref. symbol meanings: 1--\citep{1981MNRAS.195..387M};2--\citep{1993A&AS...98...51H};3--\citep{1996A&A...306..267S};4--\citep{2002ApJ...566..931S}}

\end{deluxetable}

\begin{deluxetable}{c c c c c c}
\tabletypesize{\scriptsize}
\tablewidth{0pt}
\tablecaption{H$_2$CO parameters\label{table:h2co}}

\tablehead
{
\colhead{Number}                    &
\colhead{Source}                    &
\colhead{V$_{LSR}$}                  &
\colhead{$\Delta V$}        &
\colhead{T$_{MB}$}                 &
\colhead{N[H$_2$CO]}  \\

\colhead{}                    &
\colhead{}                    &
\colhead{(\kps)}                  &
\colhead{(\kps)}        &
\colhead{(K)}                 &
\colhead{($10^{13}cm^{-2}$)}  \\
}

\startdata
001     &   00338+6312           &      -17.8(0.1)      &    2.5(0.2)         &   -0.17(0.02)    &    4.25                          \\
003     &   01194+6136           &      -14.1(0.1)      &    3.2(0.4)         &   -0.14(0.02)    &    3.70                          \\
004     &   02219+6152           &      -39.9(0.1)      &    4.0(0.1)         &   -0.82(0.02)    &    3.22                          \\
004     &   02219+6152    $^b$   &      -21.6(0.1)      &    1.2(0.2)         &   -0.12(0.02)    &    0.15                          \\
005     &   02230+6202           &      -43.3(0.1)      &    2.8(0.1)         &   -0.28(0.02)    &    1.62                          \\
006     &   02232+6138           &      -47.5(0.2)      &    4.8(0.6)         &   -0.09(0.02)    &    1.77                          \\
007     &   02244+6117           &      -49.6(0.1)      &    3.9(0.1)         &   -0.26(0.01)    &    5.34                          \\
008     &   02575+6017           &      -38.7(0.2)      &    1.7(0.3)         &   -0.06(0.02)    &    0.87                          \\
\enddata

\end{deluxetable}

\begin{table}
\begin{center}
\caption{H110$_{\alpha}$ parameters\label{table:rrl}}
\begin{tabular}{ccccc}
\tableline\tableline
Number    &   Source         & V$_{LSR}$         & $\Delta V$        & T$_{MB}$    \\ 
          &                  &           (\kps)  &            (\kps) &          (K)\\ \hline
002       &   00494+5617     &    -25.8          &      21.5         &   0.04      \\
005       &   02230+6202     &    -45.8          &      27.9         &   0.16      \\
006       &   02232+6138     &    -42.5          &      23.9         &   0.04      \\
007       &   02244+6117     &    -47.9          &      30.6         &   0.02      \\
016       &   03595+5110     &    -25.7          &      23.9         &   0.07      \\
017       &   04073+5102     &    -50.7          &      26.3         &   0.07      \\
\end{tabular}
\end{center}
\end{table}

\begin{figure*}[p]
\centering
\includegraphics[width=15.4cm]{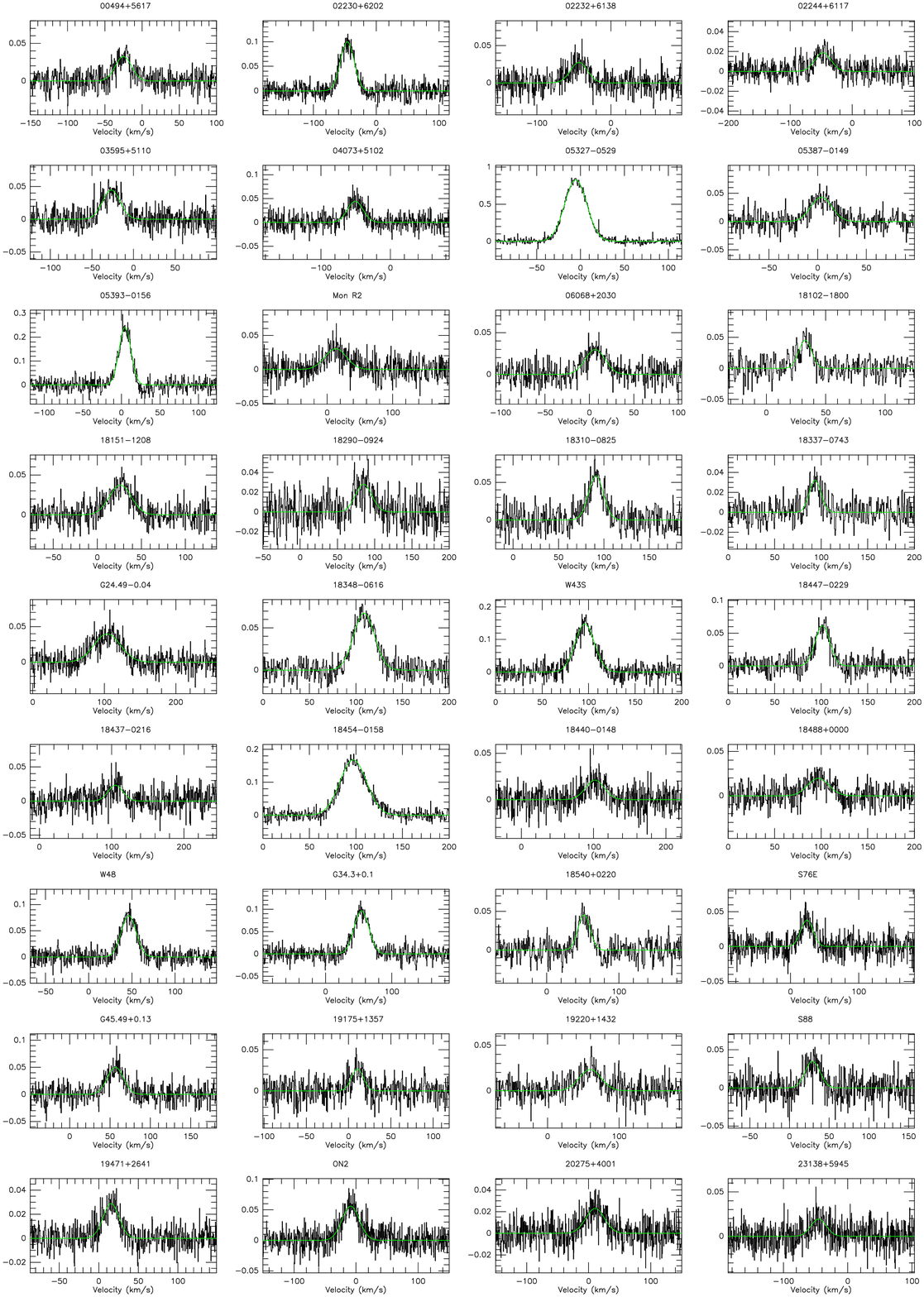}
\caption{Detected RRL lines (intensity in T$_A^*$). \label{sp1}}
\end{figure*}

\begin{figure*}
\centering
\includegraphics[width=16.4cm]{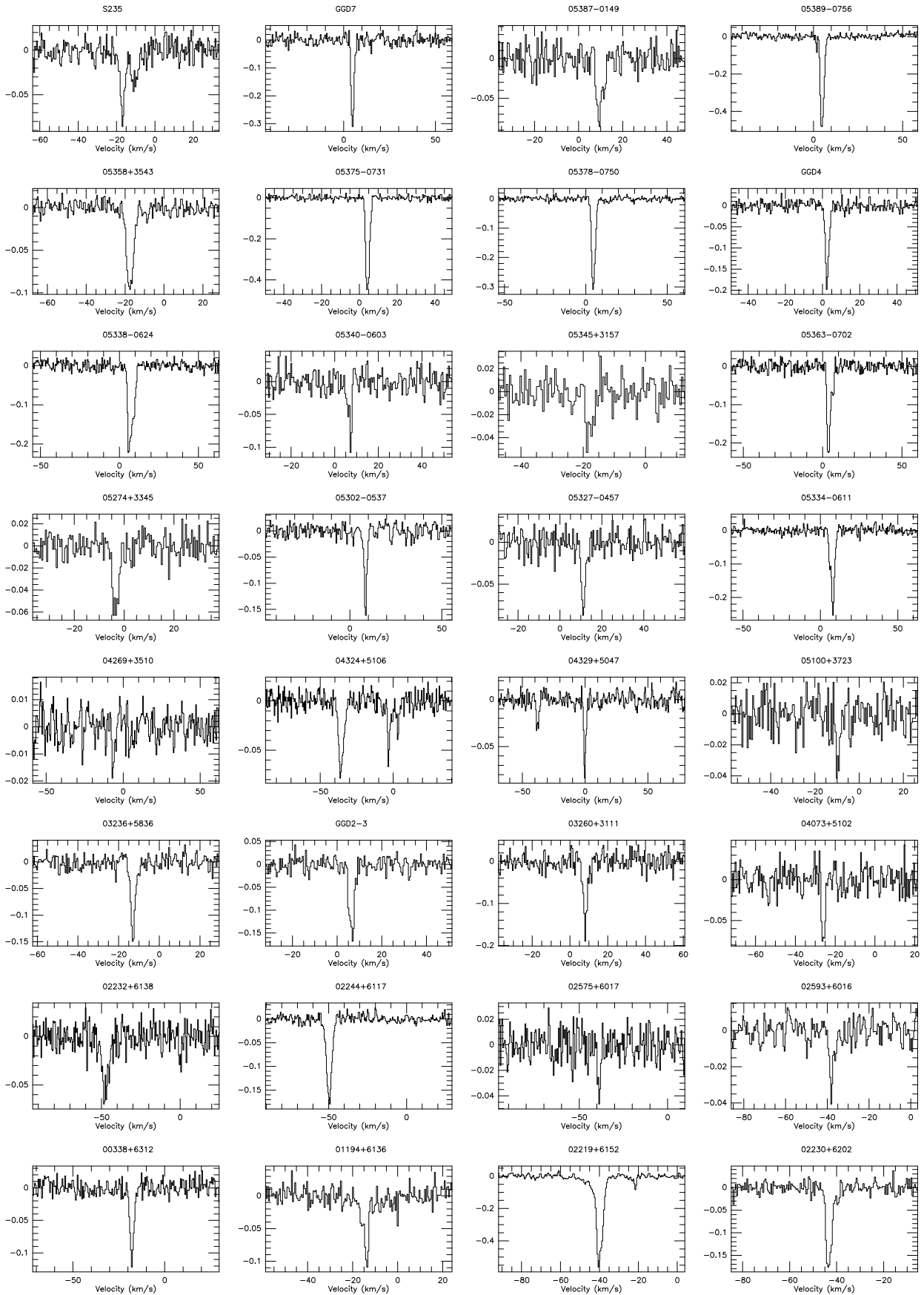}
\caption{Sources detected with H$_2$CO (intensity in T$_A^*$).\label{sp2}}
\end{figure*}

\addtocounter{figure}{-1}
\begin{figure*}
\centering
\includegraphics[width=16.4cm]{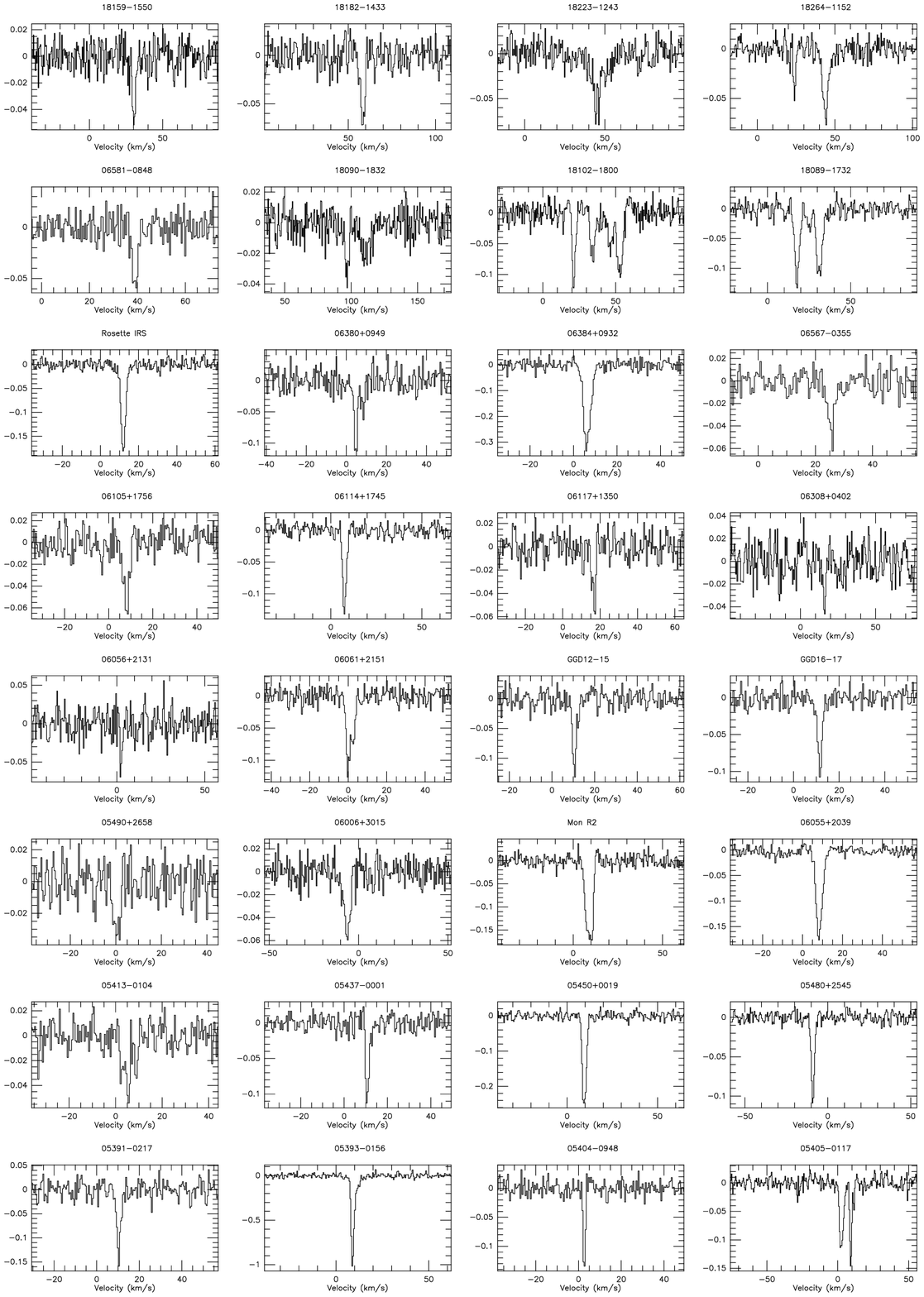}
\caption{Continued}
\end{figure*}

\addtocounter{figure}{-1}
\begin{figure*}
\centering
\includegraphics[width=16.4cm]{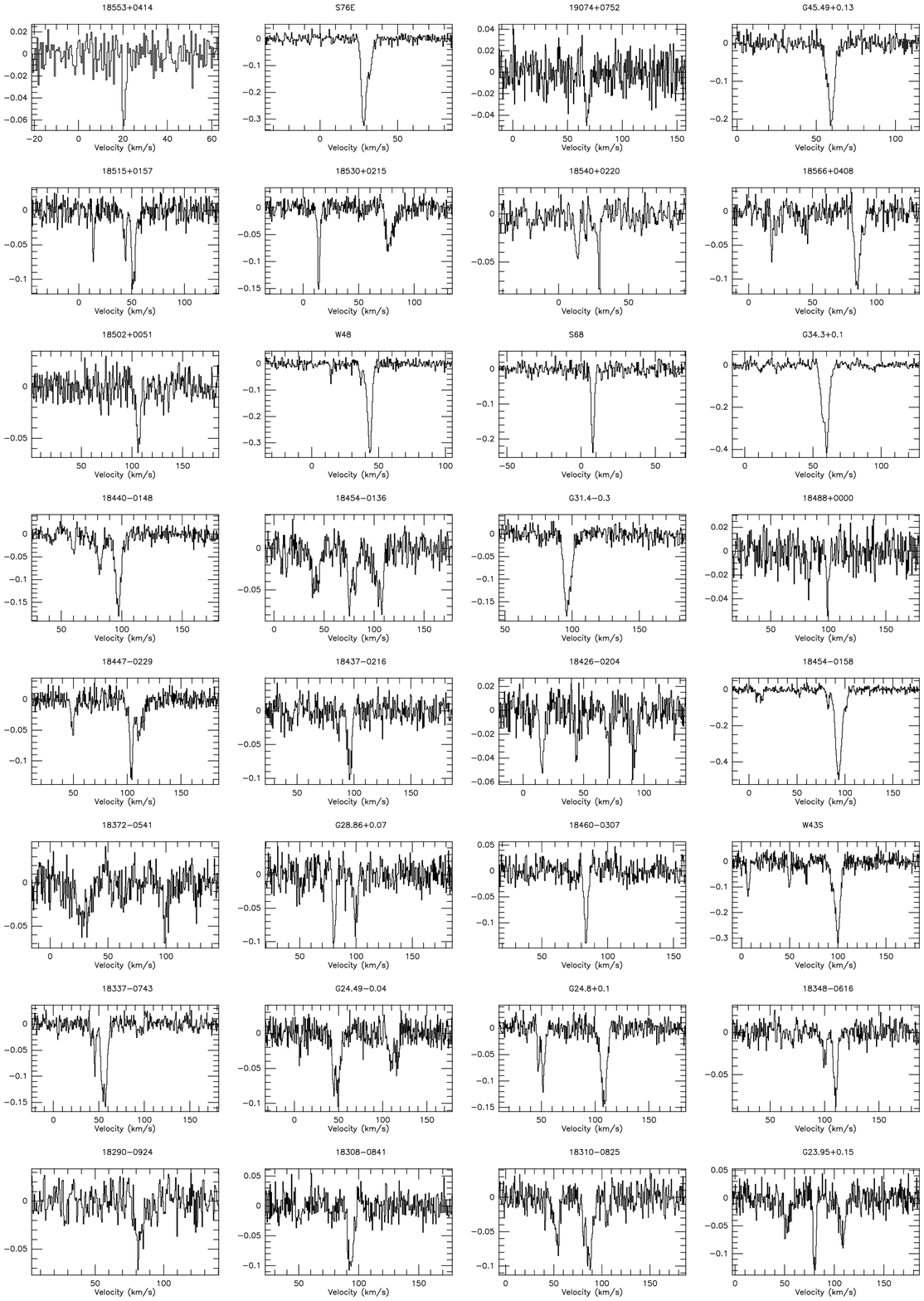}
\caption{Continued}
\end{figure*}

\addtocounter{figure}{-1}
\begin{figure*}
\centering
\includegraphics[width=16.4cm]{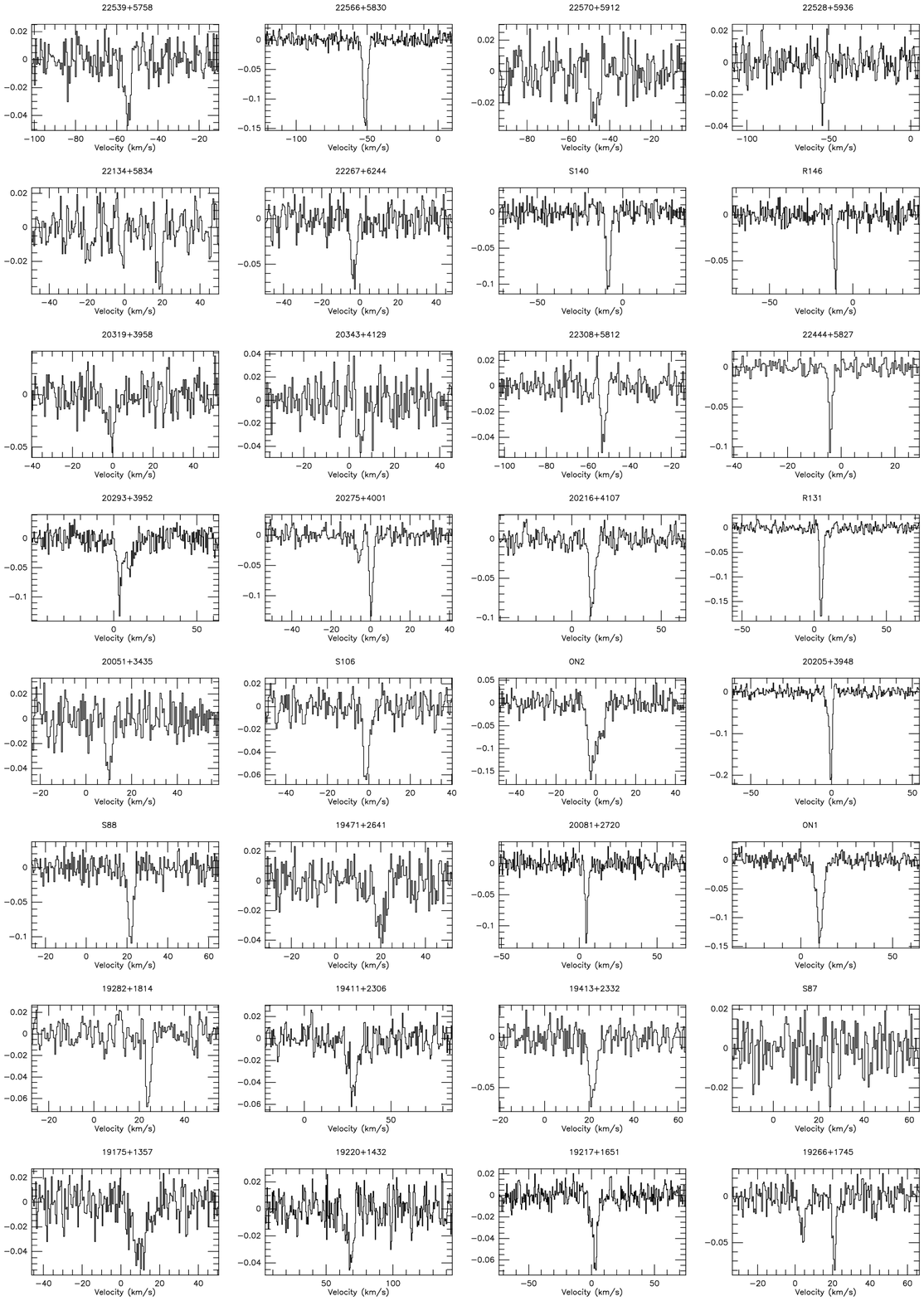}
\caption{Continued}
\end{figure*}

\addtocounter{figure}{-1}
\begin{figure*}
\centering
\includegraphics[width=16.4cm]{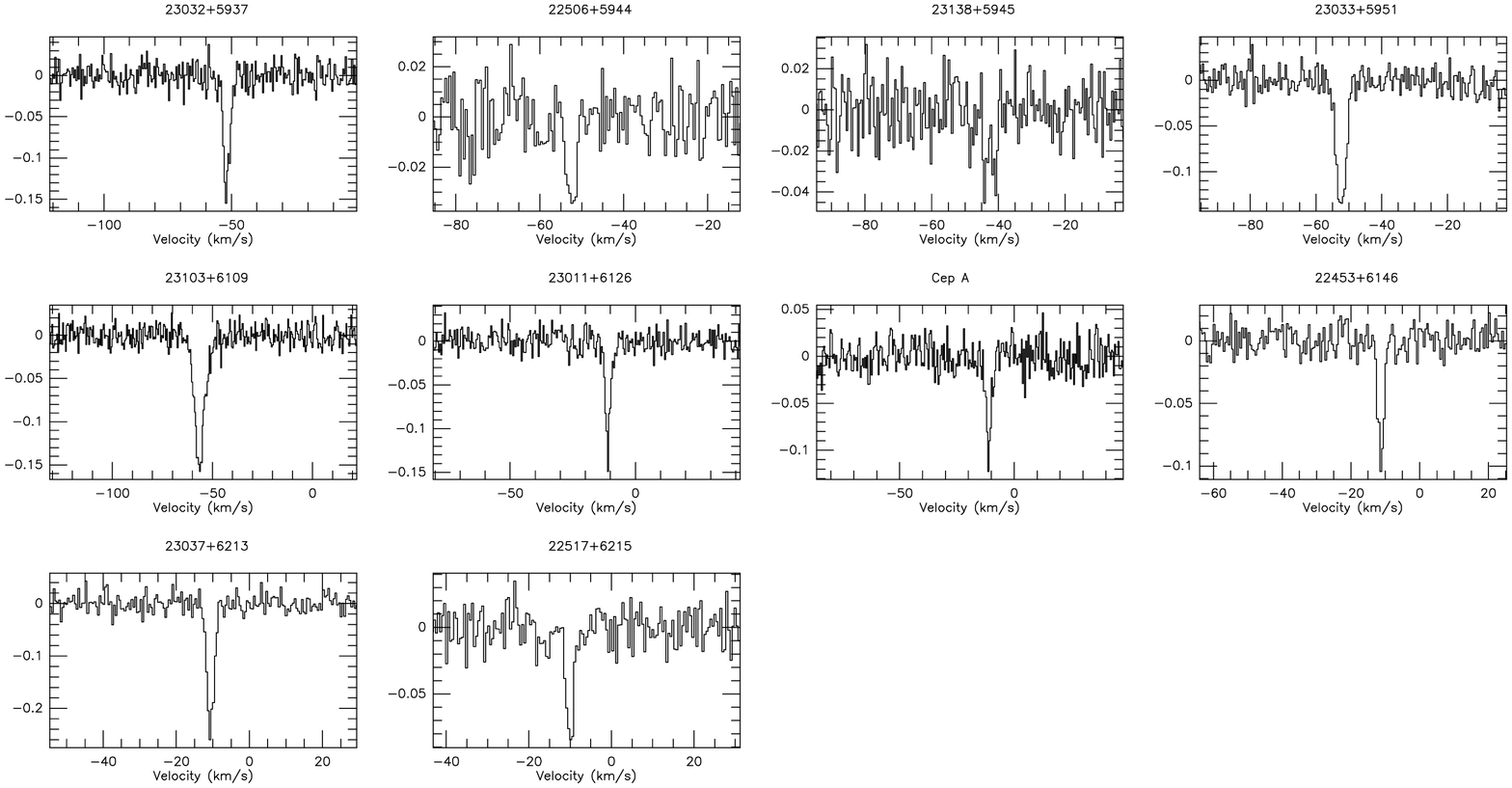}
\caption{Continued}
\end{figure*}

\begin{figure*}
\centering
\includegraphics[width=1.0\hsize,angle=90]{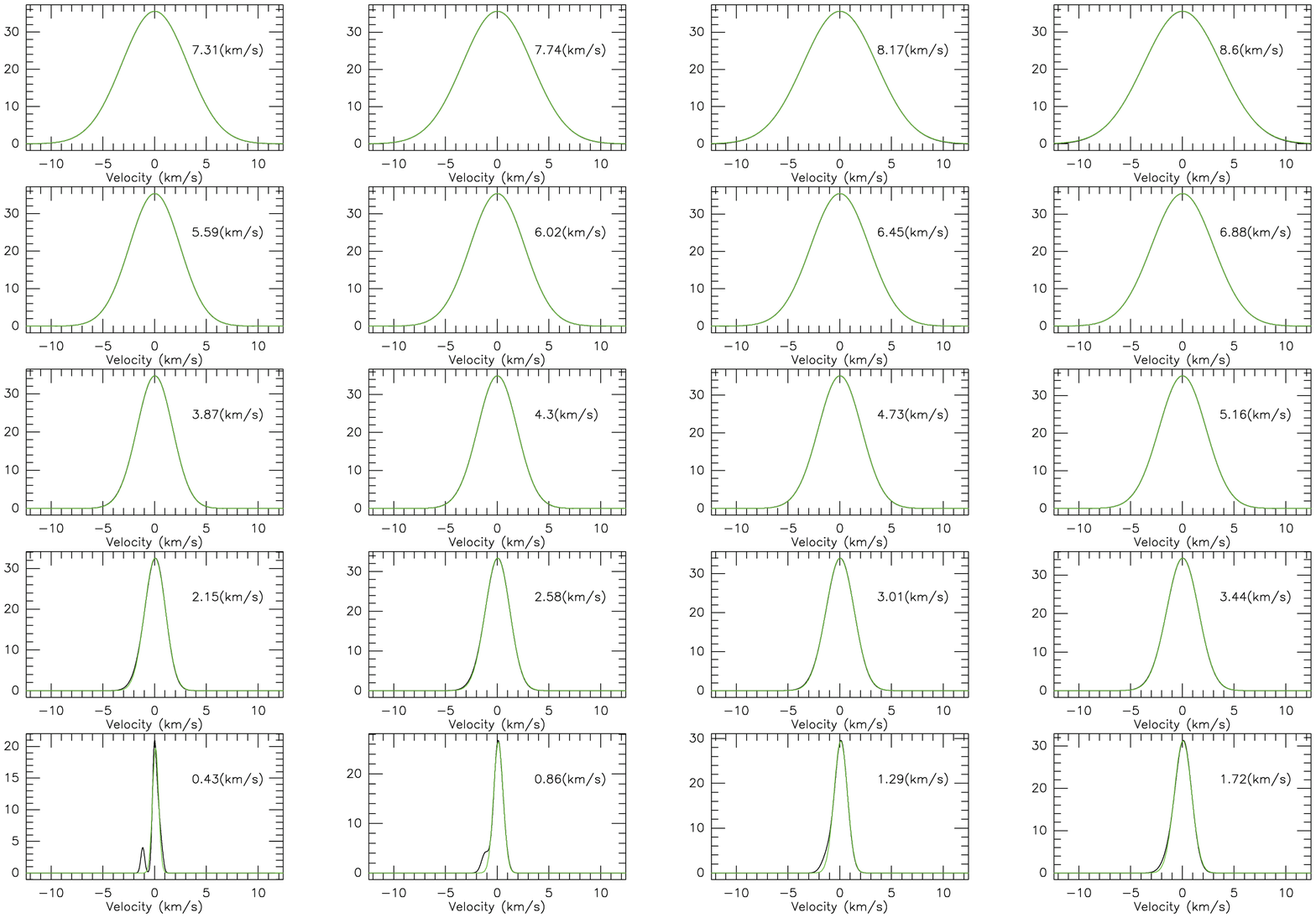}
\caption{Blended HFS simulation of H$_2$CO. The blended spectra of five HFS components are plotted in {\it black lines \rm} and the gaussian fittings are in {\it green line \rm}. }
\label{hfs2}
\end{figure*}

\end{document}